 \documentclass[tradiabstract]{aa} % for the abstract without structuration

\usepackage{graphicx}
\def\simgreat{\mathbin{\lower 3pt\hbox
     {$\rlap{\raise 5pt\hbox{$\char'076$}}\mathchar"7218$}}}
\def\simless{\mathbin{\lower 3pt\hbox
     {$\rlap{\raise 5pt\hbox{$\char'074$}}\mathchar"7218$}}}

\newcommand{\Lsun} {L$_{\odot}$}
\newcommand{\Msun} {M$_{\odot}$}

\begin{document}

\title{PAHs in protoplanetary disks: emission and X--ray destruction}

\titlerunning{PAHs in protoplanetary disks: emission and X--ray destruction}

\author {R.~Siebenmorgen\inst{1} \and F.Heymann\inst{2}}
\institute{
        European Southern Observatory, Karl-Schwarzschild-Str. 2,
        D-85748 Garching b. M\"unchen, Germany
\and
Department of Physics and Astronomy, University of Kentucky,
Lexington, KY 40506-0055, USA
}
\offprints{Ralf.Siebenmorgen@eso.org}

\date{Received February 14, 2012 / Accepted April 13, 2012}

\abstract {We study the PAH emission from protoplanetary disks.
  First, we discuss the dependence of the PAH band ratios on the
  hardness of the absorbed photons and the temperature of the stars.
  We show that the photon energy together with a varying degree of the
  PAH hydrogenation accounts for most of the observed PAH band ratios
  without the need to change the ionization degree of the molecules.
  We present an accurate treatment of stochastic heated grains in a
  vectorized three dimensional Monte Carlo dust radiative transfer
  code. The program is verified against results using ray tracing
  techniques.  Disk models are presented for T Tauri and Herbig Ae
  stars.  Particular attention is given to the photo--dissociation of
  the molecules.  We consider beside PAH destruction also the survival
  of the molecules by vertical mixing within the disk. By applying
  typical X--ray luminosities the model accounts for the low PAH
  detection probability observed in T Tauri and the high PAH detection
  statistics found in Herbig Ae disks. Spherical halos above the disks
  are considered. We show that halos reduce the observed PAH
  band--to--continuum ratios when observed at high inclination.
  Finally, mid--IR images of disks around Herbig Ae disks are
  presented. We show that they are easier to resolve when PAH emission
  dominate.}


\maketitle

\section{Introduction}

The thermal structure of the surface of protoplanetary disks, and in
particular the part which gives rise to the mid--IR emission, may be
strongly influenced by the presence of polycyclic aromatic
hydrocarbons (PAH).  Such molecules have small heat capacities and
when they absorb individual photons their temperature fluctuates
strongly.  Prominent signatures of PAHs are given by detection of
their infrared emission bands (see Tielens (2008) for a
review). Spatially resolved mid--IR observations give evidence of gaps
and spiral structures (Jayawardhana et al. 1998; Augereau et
al. 1999).  The origin of the PAH emission in protoplanetary disks is
confirmed for most of the sources (Habart et al. 2004; Lagage et
al. 2006; Doucet et al. 2007; Geers et al. 2007), although PAH
emission is observed at several thousands of AU away from the star
(Siebenmorgen et al., 1998). The detection rate of PAH bands in
protoplanetary disks for T Tauri stars is as low as 10\%, for Herbig
Ae stars as high as 60\% using ISOSWS (Acke \& van den Ancker, 2004)
and goes up to 70\% using the superb sensitivity of Spitzer (Acke et
al., 2010).  In the disks of the more luminous Herbig Be stars the PAH
detection rate is below 35\% (Acke et al., 2011).

The destruction and survival of PAH in the disks of T Tauri stars is
discussed by Siebenmorgen \& Kr\"ugel (2010, SK10).  The PAH detection
statistics in protoplanetary disks seems to be related to the
excitation mechanism of the PAHs and their destruction, in particular,
by hard photons (soft X--rays). The PAH destruction is counter--acted
by the survival of the molecules which might be caused by turbulent
mixing within the disk.  We wish to corroborate these claims by
detailed radiative transfer models of protoplanetary disks around T
Tauri and Herbig Ae stars.

This paper is structured as follows: We first study the dependence of
the strength of PAH bands on the hydrogenation coverage of the
molecules and their excitation by photons of different hardness $h\nu$
(Sect.~\ref{pah.sec}).  It is found that the spectral shape of the
exciting radiation, in particular when hard photons are involved,
should not be neglected when incorporating PAHs in radiative transfer
models. A method to include stochastic heated grains such as PAH in a
vectorized three dimensional (3D) Monte Carlo (MC) dust radiative
transfer model is presented in Sect.~\ref{3D.sec}. As a benchmark test
we compare results of the MC procedure to a ray tracing
code. Starlight heated, so called passive disks, are treated, which
are in hydrostatic equilibrium with gas and dust coupled. Disks are
irradiated by the stellar photosphere and by harder photons with
energies of several eV up to 2\,keV. Vertical mixing of the grains
within the disks is considered.  Flaring disks are computed with a one
dimensional (1D) radiative transfer code provided by Kr\"ugel (2008).
Disks are also computed with a MC code in 3D, where the combined
hydrostatic equilibrium and radiative transfer equations are solved
following the procedure by Siebenmorgen \& Heymann (2012, SH12). The
latter method was developed for disks including equilibrium heated
grains. In those disks ring--like structures and gaps appear in the
surface structure. In Sect.~\ref{disks.sec} we present results for
disks including stochastic heated dust particles.  The spectral energy
distributions (SED) are computed for pure disks and disks with dusty
halos. The surface brightness distribution in the PAH bands is
discussed. Our main findings are summarised in
Sect.~\ref{conclusion.sec}.

\section{PAH excitation and band ratios \label{pah.sec}}

Variations in the relative strength of the 11.3 $\mu$m C--H to the 6.2
and 7.7$\mu$m C=C vibrational modes of PAHs are collected for a large
sample of galactic and extragalactic objects by Galliano et
al. (2008). The authors conclude that the PAH ionisation degree
accommodates the observed variations of the band strengths (Draine
2011). We point out a degeneracy between the ionisation degree and the
hydrogenation coverage of the molecules in explaining the band
ratios. We consider a single PAH of constant: ionisation degree,
abundance (25\, ppm [C in PAH]/[H]), and size. The PAH is made up of
150 C and 37 H atoms.  The absorption cross section per C atoms in the
optical/UV is $\kappa = 7 \times 10^{-18}$\,cm$^2$ with a cut--off at
$\lambda_{\rm {cut}}= 0.6\mu$m (Schutte et al., 1993). We consider 17
emission bands and take for simplicity Lorentzian profiles. Parameters
are listed in Siebenmorgen et al. (2001).  They are calibrated using
ISO spectra of starburst nuclei. The PAH model represent an average
between neutral and ionised species. Variations of the band carriers
which may be caused by chemical and radiation processing of the
molecules in the particular environment are not treated.
Nevertheless, in disks around Herbig stars the centers of the PAH
features shift to redder wavelengths as the stellar effective
temperature decreases (Keller et al. 2008, Peeters et al. 2002). In
our models such a shift can only be compensated ad--hoc. Objects with
a strong 3.3$\mu$m band are easier fit including an additional smaller
PAH component or a PAH size distribution.  The influence of a bimodal
size distribution and the impact of the hydrogenation degree of PAHs
on their mid--IR emission is discussed by Siebenmorgen \& Kr\"ugel
(1992).  However, models including a single PAH with size as above
appears sufficient to fit SEDs of various galaxies (Siebenmorgen \&
Kr\"ugel 2007).  In order to keep computing times of the MC procedure
reasonable we treat PAHs of one size.

First we consider PAH excitation by mono--chromatic radiation and
study the influence of the hardness of the source spectrum on the band
ratios.  The emitted radiation is at constant strength (flux $F =
100$\,erg\,s$^{-1}$cm$^{-2}$), its hardness is varied using photon
energy between $h\nu = 2$\,eV and 50\,eV.  As $F \propto n_\gamma
h\nu$ is constant the number of photons $n_\gamma$ decreases as the
photon energy $h\nu$ increases. For the same strength of the radiation
field the PAH temperature distribution function $P(T)$ becomes narrow
when excited by the many soft photons and widens when excited by the
few hard photons impinging on the molecule (Fig.3 in SK10).

\begin{figure} [htb]
%\begin{center}
\hspace{-0.9cm} \includegraphics[angle=0,width=10.3cm]{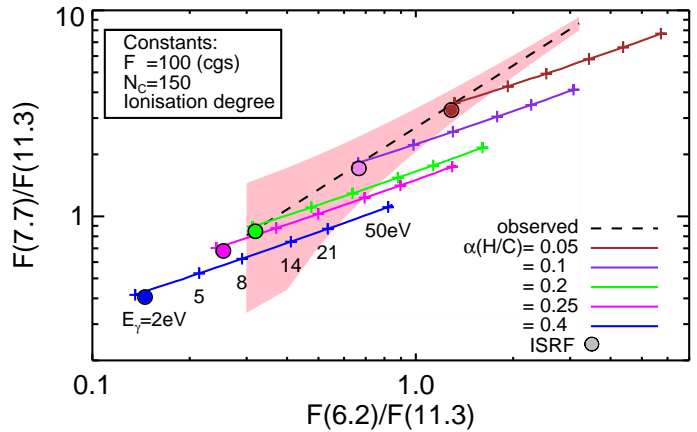}
%\end{center}
\caption{Strength of the CC stretching modes at 6.2 and 7.7\,$\mu$m
  relative to the CH out-of-plane bending mode at 11.3\, $\mu$m.
  Observed variations (shaded) and mean values (dashed) by Galliano et
  al. (2008). Full lines at various colors represent band ratios of
  PAHs with different hydrogenation, $\alpha ($H/C$)$, and excited by
  mono--energetic radiation with different hardness ($E_{\gamma}$, as
  labeled) and constant strength (flux, $F$). Band ratios derived from
  PAHs heated by the ISRF are indicated by filled circles and same
  color index of $\alpha$. Beside other parameters the degree of
  ionisation is constant. \label{pahbanden.ps}}
\end{figure}

Several teams who include PAH in MC codes (Wood et al. 2008; Baes et
al. 2011, Robitaille 2011) use pre--computed emissivity profiles of
the molecules, respectively $P(T)$. The PAH libraries are computed by
scaling a source spectrum of constant spectral shape (Draine \& Li
2007; Compiegne et al. 2011) for which the interstellar radiation
field of the Galaxy is often used (ISRF, Mathis et al., 1983). The
libraries are therefore applicable only to radiation fields in which
the strength is varied but its hardness remains close to the ISRF. In
Fig.~\ref{pahbanden.ps} we show that the band ratios of a PAH change
within a factor of three when excited by photons of different hardness
while they remain constant when excited by the ISRF. Therefore by
including PAH into radiative transfer models the frequency of the
photons exciting the molecules should be considered. This is in
particular true for hard photons of several eV well in excess of the
mean photon energy of the ISRF (1.5\,eV). In this case the
pre-computed band strengths are incorrect by a factor close to what
can be read from Fig.~\ref{pahbanden.ps}.

To cover most of the observed variations of the C--H and C=C band
ratios as second parameter, the hydrogenation of the molecule given by
$\alpha = $\,[H]/[C] is used. We vary $\alpha$ between 0.05 and 0.4
which seems within the limit of realistic H coverage of such large
molecules. We find that most of the observed values, with exception of
those at $F(7.7)/F(11.3) > 5$, are reasonably covered without having
changed the ionisation degree of the PAH (Fig.~\ref{pahbanden.ps}).

\begin{figure} [htb]
%\begin{center}
\hspace{-0.9cm} \includegraphics[angle=0,width=10.3cm]{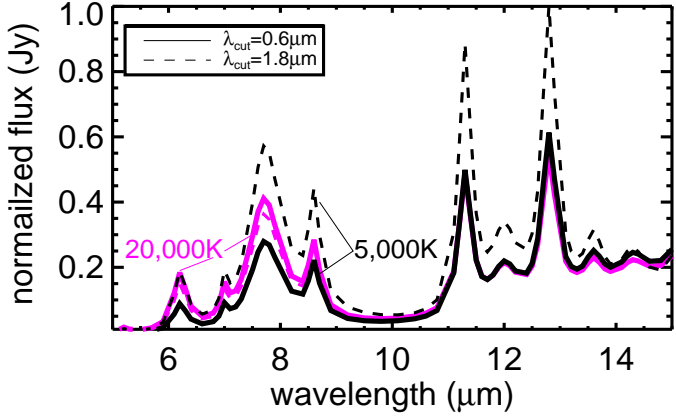}
%\end{center}
\caption{Influence of stellar temperature and cut--off wavelength of
  the PAH absorption cross section on the IR emission bands. Spectra
  are normalized to their far-IR peak. \label{sedpahbanden.ps}}
\end{figure}

\begin{table}[htb]
  
  \caption{\label{pah.tab} PAH band ratios calculated for 
    different stellar radiation fields and cut-off wavelengths 
    of the PAH absorption cross section, $\lambda_{\rm {cut}}$.}

\begin{center}
\begin{tabular}{|r| c|| c | c | c | c |}
\hline
 (1) &(2)  & (3) & (4) & (5) & (6) \\ 
 & & & &PAH &PAH \\
Star & $\lambda_{\rm {cut}}$ & $t_{\rm {abs}}$ &h$\overline{\nu}$  & ratio &  ratio \\
K    &$\mu$m & s   & eV & $6.2/11.3$ &  $7.7/11.3$ \\ 
 & & & & & \\
\hline
 5,000 & 0.6& 140& 2.6&   0.19 & 0.60 \\
20,000 & 0.6&  90& 5.3&   0.41 & 0.92 \\
\hline
 5,000 & 1.8&  25& 1.5&  0.22 & 0.66 \\
20,000 & 1.8&  75& 4.7&  0.34 & 0.83 \\
\hline
\end{tabular}
\end{center}

\hspace{0.cm} (1) Blackbody temperature (h$\nu<13.6$eV).

\hspace{0.cm} (3) Cut--off wavelength in the PAH absorption cross section.

\hspace{0.cm} (3) Average time between 2 photons absorptions (Eq.4, SK10).

\hspace{0.cm} (4) Mean photon energy absorbed by a PAH (Eq.5, SK10).

\hspace{0.cm} (5) Flux ratio at 6.2$\mu$m and 11.3$\mu$m. 

\hspace{0.cm} (6) Flux ratio at 7.7$\mu$m and 11.3$\mu$m.

\end{table}

The reflection nebulae NGC1333, which is excited by an 19,000\,K
photosphere, and NGC2068, which is heated by a 11,000\,K star, differ
in the 7.7/11.3 $\mu$m band ratio by 30\% (Uchida et al., 2000).  With
the same stellar temperatures the ratios vary in the model by 15\%,
which can be increased varying the strength of the radiation field or
the PAH parameters. The cut--off in the PAH absorption cross-section
is in the optical and moves only slightly to longer wavelength by
increasing the size of the molecule. Laboratory measurements show that
large neutral ($N_{\rm C} >30$), compact ions ($N_{\rm C} < 250$) or
nitrogenated PAHs may absorb up to the near--IR (Salama et al. 1996,
Mattioda et al. 2005, 2008). PAH emission spectra are shown in
Fig.\ref{sedpahbanden.ps} using our ISM dust model. We vary the
temperature of the star and $\lambda_{\rm {cut}}$. The stellar flux is
constant at $F = 100$\,erg\,s$^{-1}$cm$^{-2}$. Photons above 13.6eV
are truncated.  For $\lambda_{\rm {cut}}=0.6\mu$m (2\,eV) and in the
5,000\,K stellar spectrum the PAH absorb every 140\,s a photon with
mean energy of 2.6\,eV, whereas in the harder spectrum, the time
between two absorption events is reduced to $t_{\rm {abs}} = 90$\,s
and the mean photon energy equals $h\overline{\nu} = 5.3$\,eV.  The
7.7/11.3 band ratio changes by $\simgreat 50$\% for both stellar
temperatures (Table~\ref{pah.tab}), so that PAHs appear hotter in
hotter stars.  The picture changes when $\lambda_{\rm {cut}}$ is
increased to $1.8\mu$m (0.7\,eV).  Then $t_{\rm {abs}}$ and
$h\overline{\nu}$ changes as detailed in Table~\ref{pah.tab} and the
band ratios vary with stellar temperature by 25\% only.  The choice of
the PAH cross section has a strong influence on the emission spectrum.
Unfortunately the cross sections vary by large factors from molecule
to molecule.

%%%%%%%%%%%%%%%%%%%%%%%%%%%%%%

\section{Radiative transfer in 3D including PAHs \label{3D.sec}}

The radiative transfer problem is solved for arbitrary dust geometries
by a MC scheme developed by Heymann \& Siebenmorgen (2012).  It
includes grains large enough to be in thermal equilibrium with the
environment. We first give a brief description of the procedure. Then
we present an accurate method to include stochastic heated particles.
The new code is verified against computations using ray--tracing
methods.

\subsection{Large grains}

Our MC method considers advantages of different MC techniques (Lucy
1999; Bjorkman \& Wood 2001). The computational speed is increased by
calculating the flight paths of many photons simultaneously. This is
implemented by using vectorized hardware of either computer processing
units (CPU) or graphical processing units (GPU). In the procedure a
total of $N = n \cdot N_{\nu}$ photon packages are emitted per second
from a source of luminosity $L_*$. In each of the $N_{\nu}$ frequency
bins there are $n$ photon packages emitted.  Each photon package has a
constant energy $\varepsilon = L_*/N$. The model space is a three
dimensional Cartesian grid $(x,y,z)$ in which each cube can be divided
into a number of subcubes.  The flight path of each photon package is
traced through the model.  A package entering a cell interacts with
the dust if the extinction optical depth along the path within the
cell is $\Delta \tau \geq -log(\xi)$, using unified random number
$\xi$. The package is scattered if the dust albedo $A > \xi'$, using
random number $\xi'$; otherwise it is absorbed.  When the package is
scattered, it only changes direction. If the package is absorbed a new
package of same energy but different frequency $\nu'$ is emitted from
the spot of absorption.  The emission is isotropic.  Each absorption
event raises the energy of the cell by $\varepsilon$, and accordingly
its dust temperature.  In cells with extreme high optical depth photon
packages may get trapped and for computational efficiency the modified
random walk procedure by Fleck \& Canfield (1984) is implemented (Min
et al. 2009; Robitaile 2010). The SED is computed by counting photon
packets leaving the cloud towards the observer using a large
beam. Images are computed by a ray--tracer where the radiative
transfer of the line of sight from the observer, or a pixel of the
detector plane, through the model cloud is solved using the MC
computed dust absorption and scattering events.

\subsection{Small grains}

Very small grains such as PAHs show strong fluctuations of their
enthalpy state whenever they are hit by a photon. Their emission is
generally treated statistically by introducing a temperature
distribution function $P(T) dT$ which gives the chance of finding an
arbitrary particle in a temperature interval $[T, T+dT]$.  Absorption
or emission leads to a transition from an initial $U_i$ to a final
$U_f$ enthalpy state of the system, which is represented by a
transition matrix $A_{fi}$. For PAH we consider an analytic
approximation of the enthalpy $U(T)$ of graphite given by Chase et
al. (1985). The enthalpy bins are of width $\Delta U(T)$ and binned in
constant temperature intervals. Let for a given cell $l$, a number of
$n_{l, \nu}$ mono--chromatic photon packets of frequency $\nu$ be
absorbed by a PAH, then the transition matrix can be computed

\begin{equation} 
\label{afi.eq}
A_{fi} = 
\left\{ \begin{array}{ccl} 
\displaystyle{K_\nu F \over  n_{l, \nu} \ h\nu }  & : &\ \mbox{if} \ |U_f-U_i- n_{l, \nu} \ h\nu| \le
{1\over 2} \Delta U_f  \\
0   & : & \ \mbox{else} \nonumber   \end{array}  
\right. \/ 
\end{equation} 

where $K_\nu$ denotes the PAH absorption cross section.  In steady
state the sum over all k--enthalpy states $\sum A_{jk} P_k =0$, and
$P(T)$ is computed with the help of the rapid recursion formulae by
Guhathakurta \& Draine~(1989). Examples of $P(T)$ for mono--energetic
excited PAHs are discussed by SK10.  In some circumstances the
distribution function $P(T)$ is rather flat while in others it
approaches a $\delta$-function. Therefore we implement an iteration
procedure in which the temperature grid of $U(T)$ is carefully adapted
to the specific case. This ensures that numerical inconsistencies in
particular of the energy balance are avoided. Note that to reduce
computing time such an iteration procedure is not included in the MC
treatment by Bianchi (2008), who computes $P(T)$ on a fixed
temperature grid.

Stochastic heated grains have been included in MC radiative transfer
models with different levels of simplifications. Kr\"ugel (2008)
describe two ways to treat stochastic heated particles in MC schemes.
We follow the one where PAH are only excited by photons which are
directly emitted or scattered from the source. A similar approach is
taken by (Gordon et al. 2001; Misselt et al. 2001; Dullemond et
al. 2007).  A library approach is considered by Juvela \& Padoan
(2003), Wood et al. (2008), Baes et al. (2011), and Robitaille (2011),
in which $P(T)$ is pre--computed for some radiation fields such as a
scaled version of the ISRF. This method reduces the high
computational demand required to solve $P(T)$ case by case, however at
the expense of a possible computational error in particular for the
derived PAH band strengths as discussed in Sect.~\ref{pah.sec}.

Previous MC codes which include PAHs do not include the physics of the
dissociation process of the molecules; PAH photo--destruction by a
single hard photon is neglected and soft X--rays are not
considered. In our scheme we include these effects as described below.
Our code is optimised so that photons which are absorbed and
re--emitted by dust are assumed not to heat the PAHs.  This is a valid
approximation as dust emits photons in the IR which are hardly
absorbed by a PAH. This allows us to reduce storage and run time
requirements.

PAH are implemented in our code as follows: First, the source emits N
photon packages and an equilibrium of the MC model is computed by
considering the absorption but ignoring the emission of the PAHs. At
this point only the number and frequency of the packets which are
absorbed by a PAH are saved. Thereafter each cell with
$n^{\rm{PAH}}_{i}(\nu) > 0$ becomes an independent source and we
assume that it emits from its center the same number $n$ of photons at
frequency $\nu'$ given by:

\begin{equation}
  \int_0^{\nu'}  \epsilon_n(\nu) d\nu \ge \xi \int \epsilon_n(\nu) d\nu \
\end{equation}

\noindent 
with random number $\xi$ and PAH emissivity 

\begin{equation}
\epsilon_n(\nu) = K^{\rm {PAH}}_\nu \int B_\nu(T) P_n(T) dT \
\end{equation}

\noindent 
where $K^{\rm {PAH}}_\nu\ $ is the PAH absorption cross section.
% and $P_n(T)$ is solved with the help of Eq.\ref{afi.eq}.

One extra complication occurs for cells where PAH dissociate.  For
nanometer-sized grains such as PAH we consider photo--destruction. It
is estimated that a PAH with $N_{\rm C}$ carbon atoms is destroyed
(Omont, 1986), when the absorbed energy $\Delta E$ exceeds a critical
energy $E_{\rm {crit}}$\,:

\begin{equation}
\Delta E \geq E_{\rm {crit}} = \frac{N_{\rm C}}{2} \/ \ {\rm {(eV)}}
\label{pahdestruct}
\end{equation}

\noindent The energy $\Delta E$, may be either due to absorption of i)
many soft photons or ii) whenever the molecule is hit by a single hard
photon.  The first condition is only relevant at short distances from
the source while the second acts irrespective of that distance. If
Eq.\ref{pahdestruct} is fulfilled and PAHs of cell $i$ dissociate the
$n^{\rm{PAH}}_{i}(\nu)$ absorbed photon packets are again emitted from
the star.  We further simplify the method and ignore PAH absorption
during the flight path of these photons through the cloud.

\begin{figure} [htb]
\hspace{-0.7cm} \includegraphics[angle=0,width=10.1cm]{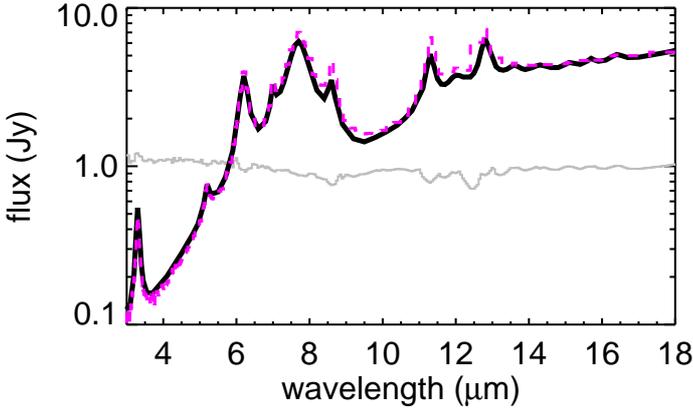}
\caption{Comparison of the SED of a stellar heated, optically thick
  ($\tau_{\rm V} =10$) dust sphere of constant density computed with a
  ray tracing method (Kr\"ugel 2008, full line) and the MC code of
  {\it {this work}} (dashed). The flux ratio between both methods is
  displayed in gray.  \label{MCHii.ps}}
\end{figure}

\subsection{Verification}

For equilibrium heated grains the MC code is verified against various
one and two dimensional benchmarks (Heymann \& Siebenmorgen, 2012).
For stochastic heated particles we compare results computed by MC
against the spheric symmetrical radiative transfer codes by
Siebenmorgen (1993) and Kr\"ugel (2008). In the latter the radiative
transfer problem is solved by a ray tracing method. We find that the
SED computed by the MC procedure and the ray tracing code agrees
typically to within a few percent ($\simless 5$\%). One such
comparison is shown in Fig.~\ref{MCHii.ps}. The dust sphere is heated
by a star with luminosity of 1\,\Lsun \/ and photospheric temperature
of 6,000\,K, between $r_{\rm {in}} = 3$\,AU and $r_{\rm{out}} =
90$\,AU a constant dust density of $1.85 \times 10^{-19}$\, g/cm$^3$
is used.  For the large grains we take ISM type dust with parameters
as in SH12 and PAHs as of Sect.\ref{pah.sec}.  This gives a dust
extinction cross section in the V band of $K^{ext}_V =
43,000$\,[cm$^2$/g--dust].  The dust mass of the cloud is $9.53 \times
10^{-7}$\Msun\/ and the optical depth between $r_{\rm{in}}$ and
$r_{\rm{out}}$ is $\tau_{\rm V} = 10$. The overall SED, including the
far--IR emission and the PAH feature strengths of both radiative
transfer computations are in good agreement (Fig.~\ref{MCHii.ps}).

\section{Protoplanetary disk models \label{disks.sec}}

\begin{table*}[htb]
\caption{\label{para.tab} Parameters of the fiducial T Tauri and Herbig Ae disks.}
\begin{center}
\begin{tabular}{|l  l | l | l|}
\hline
Parameter &  & T Tauri  &Herbig Ae  \\ 
 & & & \\
\hline
Stellar luminosity       &$L_*$ [\Lsun]      & 2      & 50  \\ 
Stellar mass             &$M_*$ [\Msun]      & 1      & 2.5  \\ 
Photospheric temperature &$T_*$ [K]          & 4,000  & 10,000\\
Column density & 
$\Sigma(r) = \frac{\tau_{\perp}(\rm{1AU})}{K_{\rm{V}}} (\frac{r}{\rm{AU}})^{\gamma}$
\,[cm$^2$/g-dust] & \multicolumn{2}{c|}{$r < 1$\,AU: \quad $\gamma = 0.5$} \\
& &  \multicolumn{2}{c|}{$r \geq 1$\,AU: \quad $\gamma = -1$} \\
Vertical optical depth & $\tau_{\perp}(\rm{1AU})$  & \multicolumn{2}{c|}{10,000}\\
Dust density in halo & $\rho_{\rm{halo}}$\,[g-dust/cm$^3$] & \multicolumn{2}{c|}{$0$ or $1.5 \times 10^{-18}$} \\
Distance                 &$D$   [pc]    & \multicolumn{2}{c|}{50} \\
Inner disk radius        &$r_{\rm{in}}$ & \multicolumn{2}{c|}{evaporation} \\
Outer disk radius        &$r_{\rm{out}}$ [AU] &  22.5 & 40 \\
\hline
\end{tabular}
\end{center}
\end{table*}

Protoplanetary disk models are set up similarly to those described in
SH12. We consider both photospheric heating and hard radiation
components of the star. In addition to large grains we include
PAHs. First, the set up of the hydrostatic disk models are described.
Then our treatment of the PAH destruction in the disks is outlined and
results compared for disks computed in 1D for flaring disks and in 2D
by the MC scheme. Models are applied to disks heated by a T Tauri star
and a Herbig Ae star. Halos above the disks are considered. The
surface brightness in the PAH bands is discussed.

\subsection{MC disk models}

As fiducial model of the T Tau star we take a mass of 1\Msun\/, a
luminosity of $L_* = 2$\Lsun \/, and a photospheric temperature of
4,000\,K.  For Herbig Ae star we use 2.5\Msun \/, 50\Lsun \/ and
10$^4$\,K (van den Ancker et al. 1997). Soft X--rays (50\,eV --
2\,keV) are included with luminosities up to $10^{-4} \ L_*$. The
distance is set to 50\,pc.

For large grains we take the fluffy grain model of SH12.  It includes
a power--law size distribution: $n(a) \propto a^{-3.5}$, with particle
radii between $320\rm{\AA} \ \leq a \leq 33\mu$m of composite grains
with volume fraction of 34\% silicates, 16\% carbon, and 50\%
vacuum. Dust abundances (ppm) are 31 [Si]/[H] and 200 [C]/[H] (Asplund
et al., 2009).  Optical constants are by Draine (2003) for silicates
and Zubko et al. (1996) for carbon. Absorption and scattering
cross-sections and the scattering asymmetry factor is computed with
the Bruggeman rule. This gives a total mass extinction cross section
for the large grains in the optical (0.55$\mu$m) of $K^{ext}_V =
4,000$\,[cm$^2$/g--dust] and a gas--to--dust mass ratio of 130. PAH
parameters are taken from Sect~\ref{pah.sec}.  Kinetic energy losses
by e$^-$ are important when dust interacts with energetic photons of
critical energy $\simgreat 100$\,eV.  This effect is not included in
the Mie theory and will result in an extra $\sim 1/\nu$ decline of the
absorption cross section; for detailed computations and size dependence 
of the critical energy see Dwek \& Smith (1996).  This correction is
applied to the cross sections of all particles.

The disk extends inwards up to the point where porous grains reach
1,000\,K, or equivalently small interstellar grains would be at about
1,500\,K. PAH destruction is computed as discussed in
Sect.~\ref{pahX.sec}. The outer disk radius is $r_{\rm {out}} =
22.5$\,AU for T Tau and 40\,AU for Herbig Ae disks. Further out the
midplane temperature drop below 30\,K and other heating mechanisms may
become important. The PAH emission is well constrained within a
fraction of $r_{\rm {out}}$. Initially we assume that the disk is
isothermal in the vertical direction $z$ and the density is given by

\begin{equation}
 \rho(r,z) = \sqrt{\frac{2}{\pi}} \ \frac{\Sigma}{H} \ e^{- z^2/2 H^2}
\label{rhothermal}
\end{equation}

\noindent
with scale height $H^2 = k T_{\rm {mid}} r^3 /G M_*m$, surface density
$\Sigma(r)$, molecular mass $m = 2.3 m_p$ and midplane temperature,
$T_{\rm {mid}}$, for which we use a power--law as an initial guess.
The height of the disk is set initially to $z_0 = 4.5 H$.  In models
with pure disks the density $\rho_{\rm{halo}}$ above $z > z_o$ is zero
and constant when a halo is considered (Table~\ref{para.tab}). The
surface density is adjusted to the optical depth in vertical
direction: $\tau_{\perp} = \Sigma(r) K_{\rm V} = \tau_{\rm {1AU}} \
(r/{\rm {AU}})^{\gamma}$ with $\gamma =-1$ for $ r \geq 1AU$ and
$\gamma =0.5$ otherwise (Min et al., 2011). At 1\,AU the vertical
optical depth from the surface to the midplane is $\tau_{\perp}{\rm
  {(1AU)}} = 10,000$. This translates to a surface density of
$\Sigma(\rm{1AU}) = 5$\,g-dust/cm$^2$ for disks without PAHs, which is
close to estimates for the minimum mass of the early solar nebulae
(Hayashi, 1981), and $\Sigma$ drops by a factor 4.4 when PAHs are
considered.

The disks are in hydrostatic equilibrium in the vertical direction so
that gravitational force is balanced by the pressure gradient
 
\begin{equation}
- \frac{z}{r} \frac{GM_*}{r^2} = \frac{1}{\rho}\frac{dP}{dz}
\label{hydro}
\end{equation}

\noindent 
with pressure $P = \rho k T(z) / m$.  We solve Eq.~\ref{rhothermal}
and Eq.~\ref{hydro} by an iterative scheme: After a first set up of
the disk structure $\rho(r,z)$ by Eq.~\ref{rhothermal}, the
temperatures $T(r,z)$ are computed with the MC code.  By inserting the
temperatures into Eq.~\ref{hydro} a new density structure $\rho(r,z)$
is found. Then with a new MC run $T(r,z)$ are updated and the
procedure repeated. After less than a dozen of iterations a stable
disk configuration is found.

\subsection{PAHs destruction and X-rays \label{pahX.sec}}

Young low-mass stars, and to a lesser degree intermediate-mass stars,
are X-ray emitters. Here we study the influence of hard photons on the
appearance of PAH bands in disks. For the destruction of PAHs the
process known as unimolecular dissociation is applied (Tielens,
2005). A simplified scheme which allows estimating of the distance
from the star where the molecules become photo--stable is described in
SK10. The threshold energy $\Delta E$ (Eq.~\ref{pahdestruct}) to
dissociate PAHs by multi-photon absorptions is given by:

\begin{equation}\label{destru}
\Delta E = N_c \int {L_\nu e^{-\tau_\nu} \ \kappa_\nu \over 4\pi r^{2}} 
  \, d\nu \ \geq E_{\rm {crit}} 
\end{equation}

\noindent 
where $L_\nu$ is the frequency dependent luminosity of the star,
$\kappa_\nu$ is the absorption cross section per C atom of a PAH and
$\tau_\nu$ is the optical depth from the star to the spot of
absorption in the cell. Hard photons can dissociate PAHs at all
distances from the source in a time short compared to the life time of
the disk. Therefore a PAH survival channel is introduced for which, as
a result of turbulence, vertical mixing of the dust within the disk is
assumed.  From Spitzer spectroscopy Keller et al. (20008) suggest that
PAH may be formed by evaporation of icy grain mantles and constantly
destroyed and replenished. We crudely simplify the picture and assume
that for the size of the largest eddies the average turbulent velocity
grows linearly with the sound speed. The turbulent velocity is
parametrized $v_{\perp} = \sqrt{\alpha} \ c_s$ with sound speed $c_s$.
Estimated values for $\alpha$ are often used but highly
uncertain. They range from $\alpha \sim 0.0001$ up to 0.1 (Schr\"apler
\& Henning 2004, Youdin \& Lithwick 2007), if not otherwise mentioned
we use $\alpha \sim 10^{-3}$. The influence of the choice of $\alpha$
on the PAH survival is discussed in SK10. For a more realistic
physical description of the turbulent motions the MC treatment has to
be combined with three dimensional hydrodynamical calculations
(Armitage 2011, for a review).  Excitation of PAHs by stellar photons
is to good order within the extinction layer of optical radiation and
has thickness $\ell \sim H/2$. For vertical velocity $v_{\perp} =
\ell/t_{\exp}$ a PAH is exposed to radiation by a time $t_{\rm
  {exp}}$. Using the destruction time $t_{\rm {des}}$, as the time
between two hard photon absorption events, the PAH will survive if
$t_{\rm {des}} > t_{\rm {exp}}$. Translated into a velocity condition
the PAH hit by photons with $E_{\rm {crit}} = h \nu_{\rm {crit}}$ will
be destroyed if (Eq.23 in SK10):

\begin{equation}\label{v_senk}
v_\perp < v_{\rm{cr}}= \ell/t_{\rm{des}} = { f_{\ell} H \over 4\pi r^2}
\int\limits_{\nu_{\rm {crit}}}^\infty {L_\nu e^{-\tau_\nu} \kappa_\nu\over
  h\nu}\,d\nu 
\end{equation}

\noindent 
with $f_{\ell} = \ell/H \sim 0.5$. Both PAH destruction conditions of
Eq.~\ref{destru} and Eq.~\ref{v_senk} are straight forward to
implement into radiative transfer codes which are based on
ray--tracing. 

By implementing the hard radiation components into the MC code we
enter into numerical challenges. The equidistant energy binning of the
photospheric component is achieved with 2,000 frequencies.  However,
considering stars with non--negligible far--UV, extreme--UV and X--ray
components means increasing the frequency grid to 250,000 bins at the
expense of a run time increase by a similar factor ($\simgreat
100$). Therefore we need to simplify the treatment of PAH dissociation
by hard photons in the MC code. The energy balance of the dust disk is
dominated by the photosphere and the hard radiation components are of
small enough luminosity that they play a minor role in the overall
disk structure and the temperature of the midplane. Hard radiation
plays a major role in computing the PAH destruction zone. The far--UV
photons have a mean energy below $E_{\rm {crit}}$ and do not
contribute to the condition of Eq.~\ref{destru}. Most of the
extreme--UV photons will be absorbed by the gas located well above the
height of dust extinction layer (Ercolano et al., 2010). Therefore it
is safe to ignore far--UV and extreme--UV radiation components. The
detailed spectral variation of X--rays is not relevant for the
destruction of PAHs. This allows us to represent X--rays by a
mono--chromatic energy which we choose to be at 1\,keV. This choice
ensures that each X--ray absorption event leads to complete
destruction of the molecule. To apply the PAH destruction condition of
Eq.\ref{v_senk} for each cell only the total optical depth, $\tau_{\rm
  {1\,keV}}$ remains to be computed. Therefore the survival of PAHs in
a cell can be estimated before the time consuming MC run. In cells
where PAHs are destroyed, by the condition of Eq.~\ref{v_senk}, there
will be no photon absorption by PAHs.

\begin{figure}  [htb]
%\begin{center}
\hspace{-0.8cm} \includegraphics[angle=0,width=10.2cm]{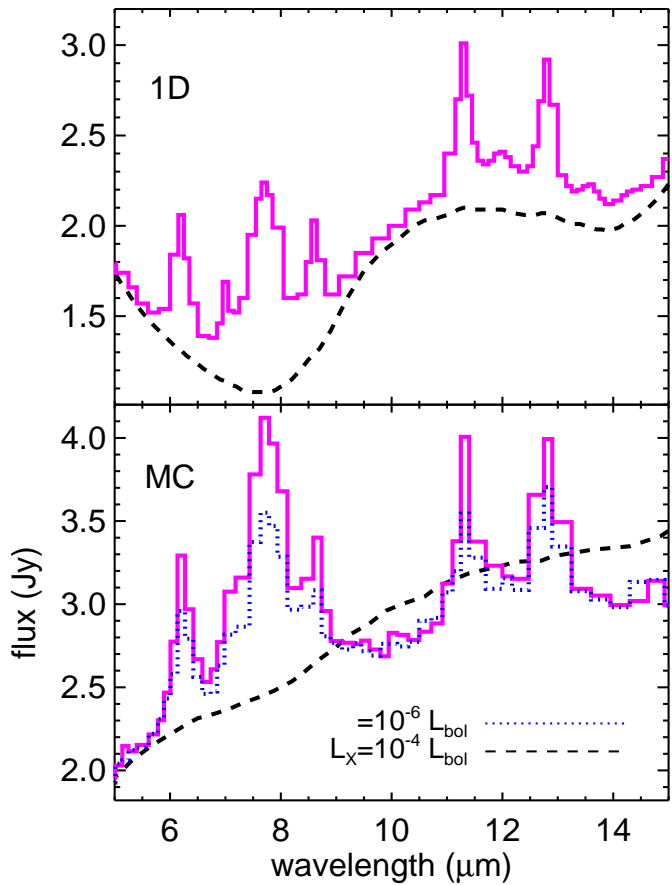}
%\end{center}
\caption{Mid IR emission of T Tau disks with X-ray luminosity of
  $L_{\rm {X}} = 10^{-4} L_*$ (dashed), $L_{\rm {X}} = 10^{-6} L_*$
  (dashed), and $L_{\rm {X}} = 0$ (full line).  SEDs are computed for
  flaring disks in '1D' ({\it {top}}) and with MC
  ({\it{bottom}}). \label{TTS_sedXdest.ps}}
\end{figure}

\begin{figure}  [htb]
%\begin{center}
\hspace{-0.8cm}
\includegraphics[angle=0,width=10.3cm]{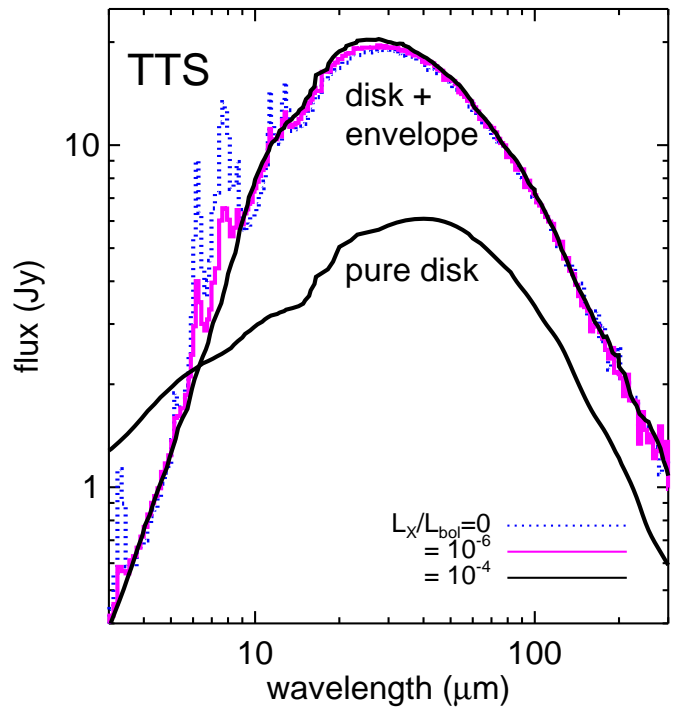}
%\end{center}
\caption{IR emission of T Tau disks with and without dust envelope,
  viewed nearly ($\sim 20^{\rm {o}}$) edge--on. Heating includes
  X--ray luminosities as indicated.
\label{TTSh_sedXdest.ps}} 
\end{figure}

\begin{figure}  [htb]
%\begin{center}
\hspace{-0.8cm}
\includegraphics[angle=0,width=10.3cm]{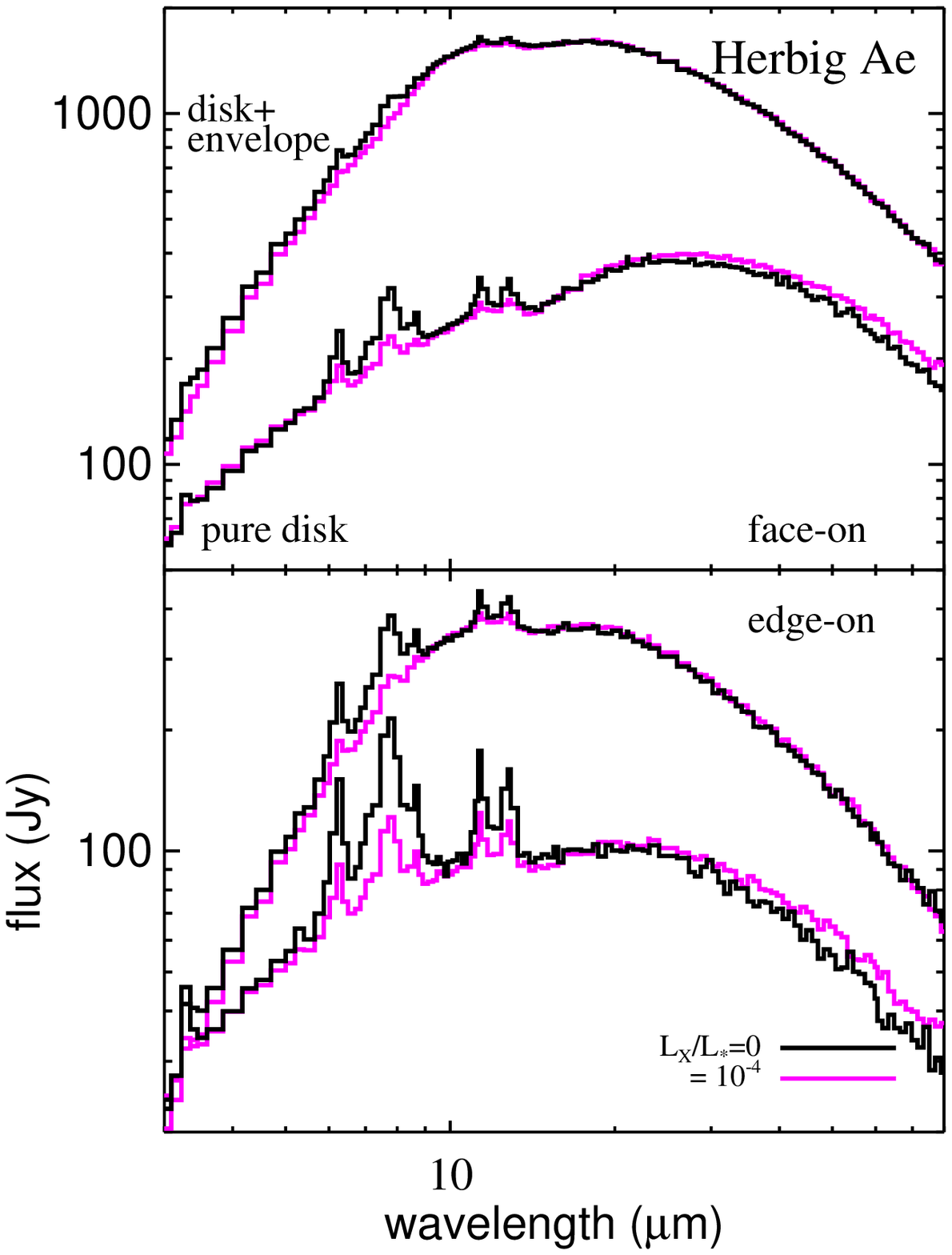}
%\end{center}
\caption{IR emission of a Herbig Ae disks with (magenta) and without
  (black) dust envelope, viewed nearly face-on ({\it top}) or edge--on
  ({\it bottom}). Results for different X--luminosities as
  indicated. \label{hebe_sedXdest.ps}}
\end{figure}

\subsection{T Tauri disks \label{tts.sec}}

We compute the SED of hydrostatic disks with PAHs and hard radiation
components for T Tauri stars applying three different disk
geometries. First, we consider a flaring disk in which the radiative
transfer is solved only in the vertical direction. Then we apply the
MC code to disks assuming axial symmetry. In the MC models the region
above the disk is either free of dust or filled in a halo like
configuration.

We apply the one dimensional (1D) radiative transfer code developed by
Kr\"ugel (2008) of a hydrostatic and geometrically thin disk. The
program is applied to the emission of PAHs from disks of Herbig stars
by Habart et al. (2004). The disk is symmetric with respect to the
midplane at $z=0$. The density structure is given in cylindrical
coordinates $\rho(r,z)$. Light from the star falls on the disk under a
grazing angle $\alpha_{\rm{gr}}$ so that from everywhere on the disk
surface the star is visible (Armitage, 2007). For the 1D models we use
the flaring geometry by Chiang \& Goldreich (1997) set to
$3^{{\rm}{o}} \leq \alpha_{\rm{gr}} \propto r^{2/7} \leq
7^{{\rm}{o}}$. We present results where the 1D disks are heated by
four radiation components of the star (SK10): the photosphere with
luminosity of 0.99\,$L_*$ and black body temperature of 4,000\,K,
far--UV with 1\%\,$L_*$ at 15,000\,K, extreme--UV with 0.1\%$\,L_*$ at
300,000\,K and X--rays of up to 2\,keV with $L_{\rm X} = 10^{-3.6}
L_*$ and a spectrum $\propto \nu^2$. We implement into the code the
PAH destruction mechanism following conditions of Eq.~\ref{destru} and
Eq.~\ref{v_senk}.

The SEDs of T Tauri disks including PAHs and heated by hard radiation
are computed also by the MC scheme for which parameters of
Table~\ref{para.tab} are taken.  We find that the appearance and
non--appearance of PAH bands is predicted in a similar way in both
geometries provided by the 1D and the MC model.  As shown in
Fig.~\ref{TTS_sedXdest.ps}, if one considers only the photosphere
strong PAH bands appear in the spectra. They weaken when including
hard radiation fields of $L_{\rm {X}} = 10^{-6} L_*$ and PAH do not
contribute to the emission spectrum for $L_{\rm {X}} \geq 10^{-4}
L_*$. The latter X--ray luminosity can be considered as a lower limit
for most T Tauri stars (Preibisch et al., 2006). Our computations
corroborate earlier estimates that the non--appearance of PAH bands in
spectra of T Tauri stars is connected to the destruction of the
molecules by hard radiation (SK10).  We verified that X--ray
luminosities have little impact on the midplane temperature, nor do
they impact the detailed surface structure of the disks. The
ring--like structures, their shadows and gaps in the disk surface are
preserved as computed by SH12.

\begin{figure*}  [htb]
\begin{center}
\hspace{-0.4cm}\includegraphics[angle=0,width=8.2cm]{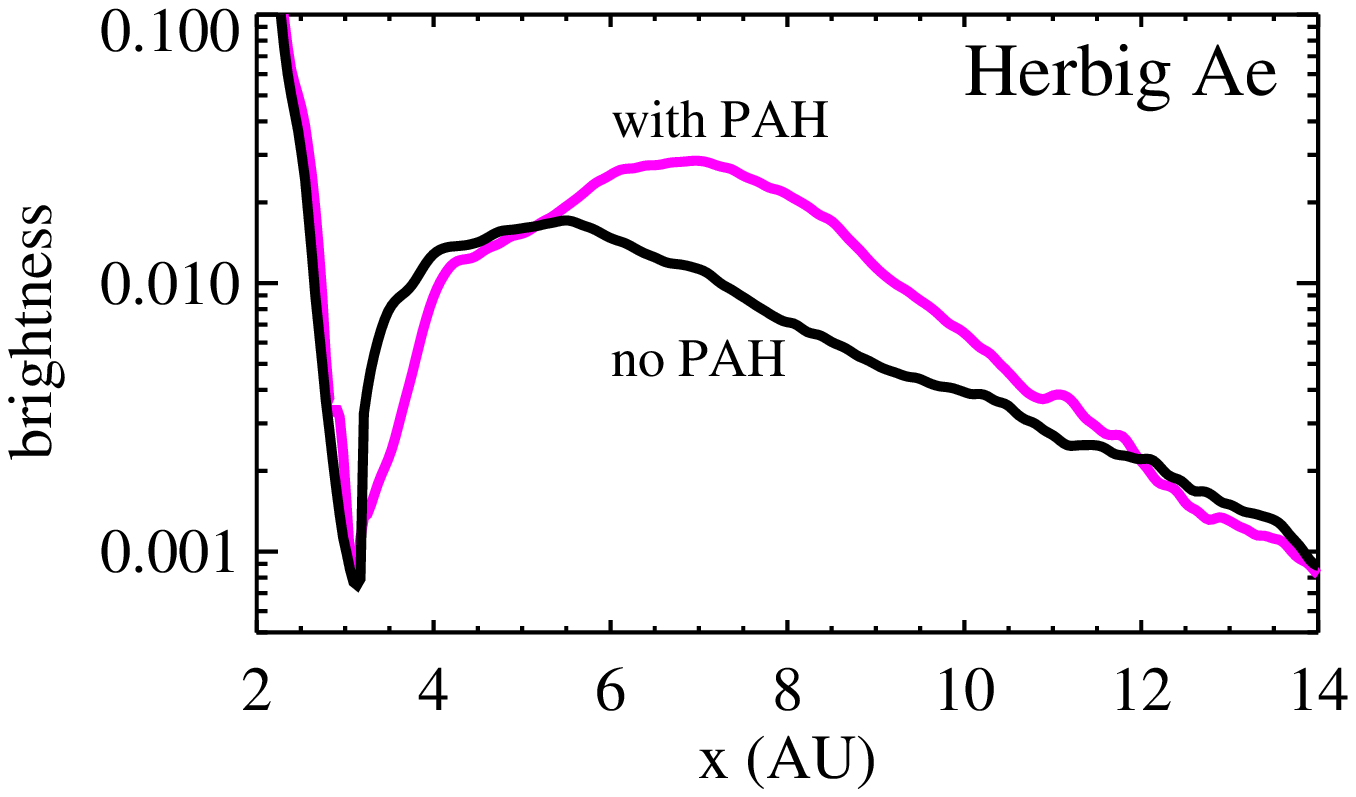}
\hspace{-0.6cm} \includegraphics[angle=0,width=5.75cm]{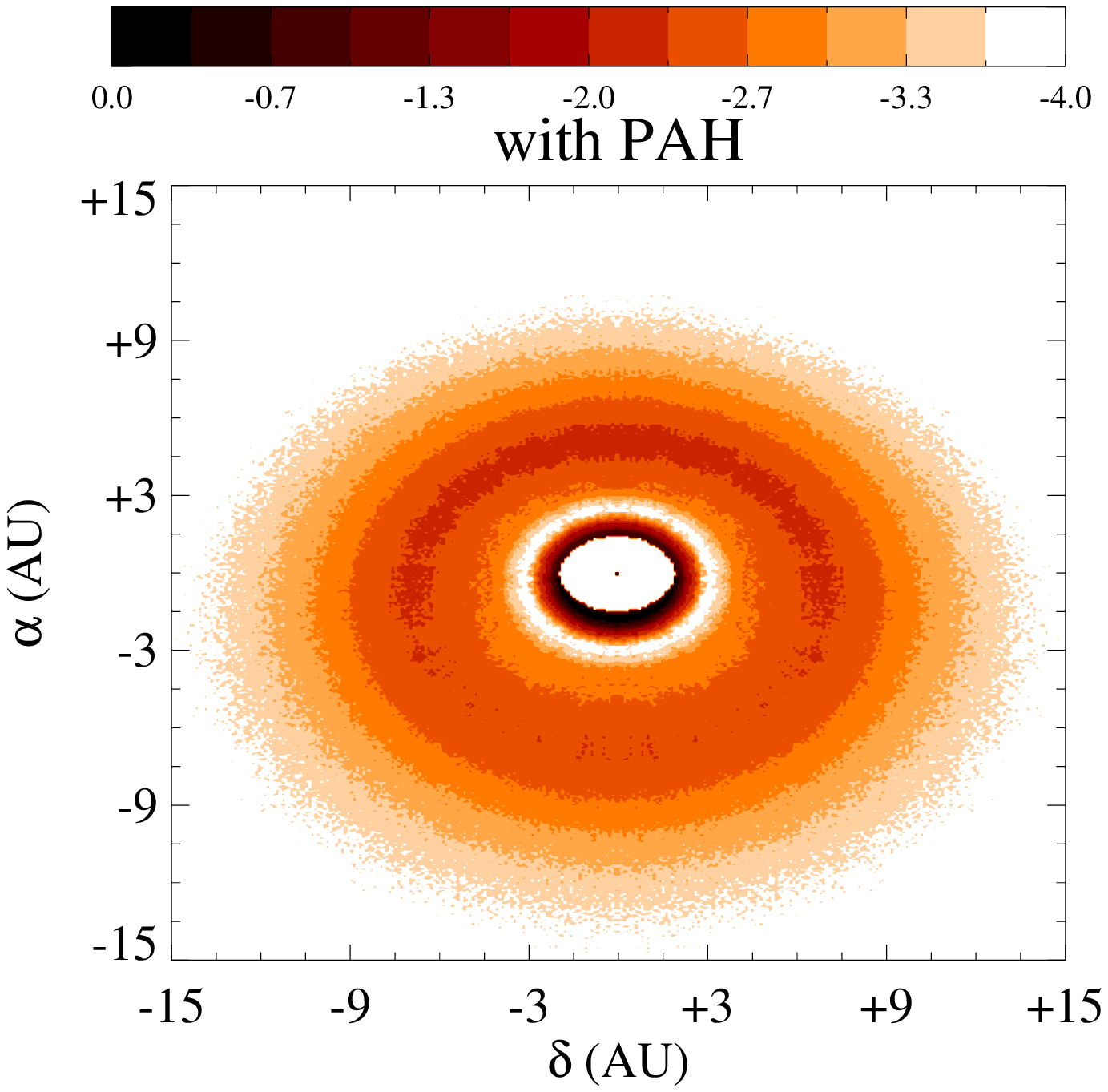}
\hspace{-0.6cm}\includegraphics[angle=0,width=5.75cm]{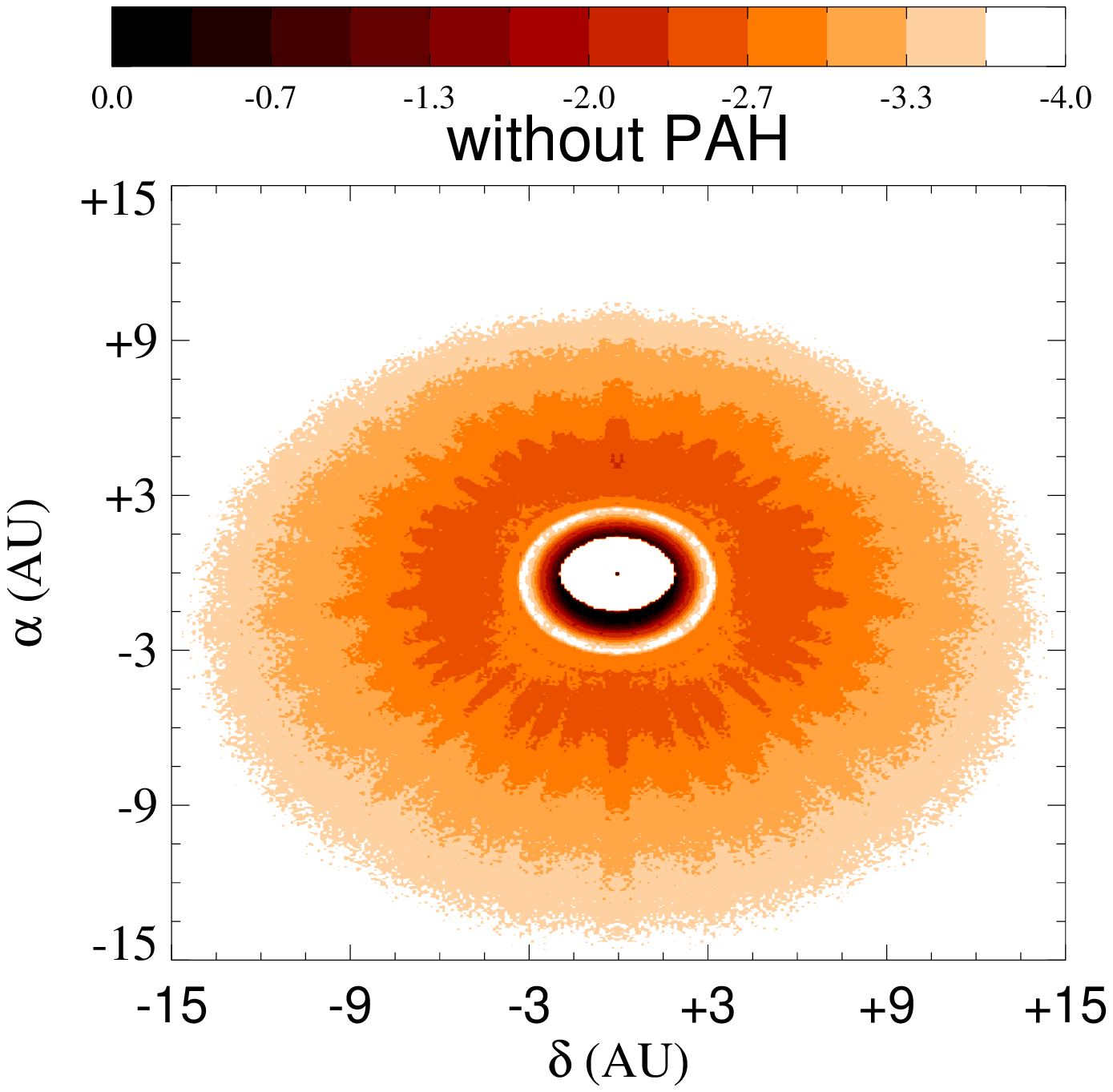}
\end{center}
\caption{Mid-IR images at 11.3$\mu$m of a disk with ({\it middle}) and
  without ({\it right}) PAHs heated by a Herbig Ae star and viewed at
  30$^o$ from the midplane. Images are normalized to the stellar
  flux. Color bars are in log(Jy/arcsec$^2$). {\it Left: } Surface
  brightness distribution of the images measured as cut from the
  star.  \label{hebe113.ps}}
\end{figure*}

\subsection{Halos}

Residual dust may be left or replenished in a halo at high altitudes
above the disk (Vinkovic et al. 2006; Krijt \& Dominik 2011).  Early
models added spherical distributed and optical thin dust emission on
top of SEDs computed for pure disks (Miroshnichenko et al., 1999). The
MC code allows a self--consistent treatment of the radiative transfer
in such a geometry. The halo is filled by large grains and PAHs and is
implemented into the code assuming a dust density of $\rho(r,z) \geq
\rho_{\rm{halo}} = 1.5 \times 10^{-18}$\,(g-dust/cm$^3$). This gives
an optical depth of the halo of $\tau_{\rm V} \sim 1.5$. Resulting
SEDs are shown in Fig.~\ref{TTSh_sedXdest.ps}.  The X--ray luminosity
is varied using $L_{\rm{X}} = 0$, $10^{-6} L_*$, and $10^{-4} L_*$.
The dependence of the PAH band strengths on the X--ray luminosity is
similar to models without halos (Fig.~\ref{TTS_sedXdest.ps}). The halo
models appear warmer and have a factor $\sim 3$ stronger IR peak
emission and at $\simless 6\mu$m less near IR flux than pure disks.
The halo provides additional heating of the disk on top of the direct
stellar light and therefore their disk midplane temperature is warmer
and shows less structure.

\subsection{Herbig Ae disks}

We present disk models for Herbig Ae stars with parameters of
Table~\ref{para.tab}. Resulting SEDs are computed by the MC scheme and
are shown for face--on and edge--on views in
Fig.~\ref{hebe_sedXdest.ps}.  The optical depth of the halo is
$\tau_{\rm V} \sim 1.5$.  The influence of hard radiation on the PAH
destruction is studied by applying a X--ray luminosity of $L_{\rm{X}}
= 0$ and $10^{-4} L_*$, which exceeds the range of $10^{-5}... \
10^{-7} L_*$ typically found in Herbig Ae stars (Stelzer et al.,
2006).  Irrespective of these X--ray luminosities PAH bands are
visible in the Herbig Ae disks. This is contrary of what we found for
T Tauri disks (Sect.~\ref{tts.sec}). At the inner rim of the disks the
absolute X--ray flux is more than a factor 10 lower in Herbig stars
than in T Tauri stars and PAH survive.  If one lowers $\alpha$ by one
order of magnitude the PAH band--to-continuum ratio is reduced by 30\%
and for $\alpha=0$ the PAH signatures are absent. We verified that the
up and downs in the midplane temperature as well as the ring--like
structures, the shadows and gaps in the disk surface are preserved as
computed by SH12, irrespectively of the X--ray luminosity or the
strengths of the PAH emission.

When the disks are viewed face--on, the PAH feature--to--continuum
ratio becomes smaller and even more so when a halo is present.  In the
latter configuration the emission by large grains outshine the PAH
contribution. We noticed a similar effect in disks of T Tauri stars in
which the X--ray luminosities are small enough for PAH survival.
Therefore, the geometrical distribution of the dust effects the
detection probability of PAH features in disks.

The stochastic heating of PAH is rather independent of the distance to
the star, a fact first noticed in reflection nebulae (Sellgren et al.,
1985).  We study how this effect influences the appearance of pure
Herbig Ae disks in the mid--IR.  The surface brightness distribution
in the 11.3$\mu$m band is displayed in Fig.~\ref{hebe113.ps} for disks
with and without PAHs. The images with PAH emission are brighter than
those without PAHs at distances between 6\,AU and 10\,AU. Otherwise,
the emission structure remains the same, except for some broadening of
the inner gap in the disk, outside the dust evaporation zone. In both
cases the dominant contribution of the emission is coming from the
central 2--3\,AU. Due to the brighter optical luminosities of Herbig
Ae stars the PAH are excited at large distances from the star. There
the turbulent velocity is larger than the critical velocity for PAH
destruction by X-rays, so that the molecules survive
(Eq.~\ref{v_senk}). For $\alpha$ close to 0 the surface brightness
distribution approaches that of disks without PAHs.

%%%%%%%%%%%%%%%%%%%%%%%%%%%%%%%%%%%%%%%%%%%%%%%%%%%%%%%%%%%%%%%%%%%

\section{Conclusion \label{conclusion.sec}}

We study the dependencies of the PAH band ratios by changing the
hydrogenation coverage of the molecules and the spectral shape of the
exciting radiation field. PAH band ratios vary by a factor three when
excited by photons of different hardness. They remain almost constant
when excited by the ISRF. Therefore by including PAH into radiative
transfer models the frequency of the absorbed photons should be
considered.

MC radiative transfer models which treat stochastic heated grains
implement different levels of simplifications: Temperature
distribution functions are often used which are pre--computed for a
constant hardness of the exciting radiation, photo--dissociation of
the PAHs is not included and hard radiation components are
ignored. The MC code presented in this paper considers heating of the
PAH by individual photons and photo--destruction of the molecules.  We
compare our PAH treatment in the MC scheme against a radiative
transfer program which is based on ray--tracing.  The PAH emission
computed by both codes is in good agreement.

The PAH emission in protoplanetary disks is studied for T Tauri and
Herbig Ae stars. Dust in the disk is configured either in a slab
geometry, in which the radiation transport is solved in vertical
direction (1D), or by the MC procedure assuming hydrostatic and
radiation balance (SH12). The disks are heated by the stellar
photosphere and soft X--rays. Hard photons dissociate PAHs in the disk
within a time that is short compared to the lifetime of the
disk. Therefore, following SK10, a PAH survival channel is introduced
in the radiative transfer programs for which, as a result of
turbulence, vertical mixing of the dust within the disk is assumed.
We find that the appearance and non--appearance of PAH bands is
predicted in a similar way by the 1D and the MC disk models.

We find that stars with X--ray luminosities $\ge 10^{−4} L_∗$ do not
display PAH emission features in disks around T Tauri stars whereas
PAH are present in disk models of Herbig Ae stars because their
absolute X--ray flux is more than a factor 10 lower. Due to the
brighter optical luminosities of Herbig Ae stars, the PAH are excited
at large distances from the star. There the turbulent velocity is
larger than the critical velocity for PAH destruction, so that the
molecules survive and PAH emission can be detected. The dependence of
the X--ray luminosity on the PAH band ratios is similar in disk with
and without halos.  However, in disks with halos the PAH band ratios
are smaller than in disks without halos.  The mid--IR surface
brightness appears brighter in the planet forming region of disks with
PAH emission than in those without PAHs.

\begin{acknowledgements} {We are grateful to Endrik~Kr\"ugel for
    helpful comments and for providing his 1D radiative transfer
    codes.}

\end{acknowledgements}

\end{document}